\documentclass[10pt]{article}
\usepackage[utf8]{inputenc}
\usepackage[T1]{fontenc}
\usepackage[margin=1in]{geometry}
\usepackage{booktabs}
\usepackage{url}
\usepackage{hyperref}
\usepackage{amsmath}
\usepackage{amssymb}
\usepackage{graphicx}
\usepackage{xcolor}
\usepackage{listings}
\usepackage{float}
\hypersetup{colorlinks=true,linkcolor=blue!50!black,citecolor=blue!50!black,urlcolor=blue!50!black}

\title{\textbf{Scan the Skill, Govern the Action:\\
Composing Registry Verdicts with Runtime Consequence Control}}
\author{
  Rohit Taneja\\ Pheo Inc\\ \texttt{rohit@pheo.ai}
  \and
  Travis Weber\\ Pheo Inc\\ \texttt{travis@pheo.ai}
}
\date{}

\begin{document}
\maketitle

\begin{abstract}
Agent skill registries screen what they publish. OpenClaw's own security team
published a careful measurement of that screening, reporting that its scanners
overlap on at most 10.4\% of combined positives and that 81.9\% of flagged skills are
caught by one scanner alone~\cite{clawhubsignals}. We take that finding as given, and
suggest the interesting question it raises is not \emph{which scanner is right} but
\emph{which question is being asked}. Each signal in that pipeline answers some form
of ``is this skill malicious?'', which is the question it is built to answer and
answers well. A second question, ``is this action permitted on this machine, by this
operator, right now?'', is not one any of them is designed to express, and it has a
different answer on the same artifact depending on whose infrastructure it runs on.

We report three measurements over 66,192 public ClawHub skill versions and a live
agent. First, 705 skills across 135 distinct publishers that all three scanners and
the registry's own judge rate clean nonetheless instruct an action many operators
forbid outright, an action named as prohibited by CIS Control 2.7 and NIST SP 800-53
CM-11. A hand audit of 100 of them puts our detector's precision at 92\% (95\% CI
[84.8\%, 96.5\%]) and found no marker of malicious intent in any correct detection.
One publisher accounts for 506 of the 705, so the finding is 135 independent parties
rather than 705 independent decisions, and we report all of these numbers rather than
the largest. Permission and maliciousness are different predicates over the same
artifact, and the pipeline computes the one it was built to compute. This measurement
is reproducible end to end from public artifacts and we give the command.

Second, of 144 commands a live agent executed while following real skill
documentation in a sandbox (39 skills, one agent), 2 appear verbatim in that
documentation, and more to the point \textbf{34.7\% had a consequence class that
appeared in no code block of the document at all}. Static analysis reasons about a
document that is an imperfect bound on what runs. Third, over 53 cleared skills that
document an action no clean record earns, the agent reached for one in 23 and the
gate stopped all 23. The harness and every command are released.

We also set out what sits outside each layer's view, including our own: a gate that
reads actions does not see harm that never becomes one. Those gaps do not coincide,
which is why we think the layers compose rather than compete.

We offer one design for a control point at that gap: a deterministic resolver
(no model in the decision path; 67.6\,ms median end to end through the current CLI
hook, almost all of it process spawn) feeding a per-(resource, class) trust
ledger whose promotion thresholds are derived from an operator's own stated risk
tolerance rather than asserted. The common flat rule of
``ten clean approvals'' cannot exclude a true failure rate as high as 25.9\% at 95\%
confidence, and demands 3--15$\times$ less evidence than the operator's own stated
policy requires. We argue
agent-skill safety is a composition of two different questions, not a contest
between two detectors, and we report where our own layer fails as precisely as
where it helps.
\end{abstract}

\section{Introduction}
\label{sec:intro}

No bank checks a customer once. Know-your-customer runs at account opening: identity
verified, source of funds established, sanctions lists cleared. It is a real control
and a necessary one. Then the account opens, and a second control takes over that
nobody would consider optional: every transfer afterwards is screened as it
happens, against limits, against patterns, against what this account has done
before. No regulator would accept ``we verified them at onboarding'' as a complete
answer, because the customer who was verified on day one is not a promise about the
transfer attempted on day four hundred.

Agent skills today have the first control and not the second. Registries screen a
skill before it is published, thoroughly, with several independent instruments, and
then an autonomous agent installs it and begins acting on a real machine with real
credentials, and no control asks whether any particular action is one this
organisation permits. This paper is about the second half of that pattern: what it
costs to be missing, and what it takes to build.

That a second control is required is settled rather than open, and a commercial tier
has formed around it (\S\ref{sec:related}). We take that as established. Our question
is narrower: given that something must sit in front of every action, on what
principled basis does it ever stop asking? A control that interrupts an operator on
every consequential action converges on being switched off, and an operator who
switches it off is less protected than one who never installed it.

An agent skill is a directory containing a \texttt{SKILL.md}: natural-language
metadata describing when a skill applies, and instructions an agent follows once it
does. ClawHub is the public registry through which most OpenClaw skills are
distributed. Before a skill version is published, it passes a pre-catalog
verification gate: independent scans from VirusTotal, in-house static analysis, and
NVIDIA SkillSpector are handed as context to ClawScan, an LLM-as-judge harness that
weighs them alongside provenance, metadata, and moderation history and issues one
verdict: Clean, Suspicious, or Malicious~\cite{openclawblog}. The dataset records
which model produced each verdict, and we report it from the data rather than from
the vendor's description: GPT-5.5 accounts for 99.47\% of the 66,192 rows, with the
remainder from \texttt{gpt-5-mini} (0.42\%) and a \texttt{codex-security-worker}
(0.10\%)~\cite{clawhubdataset}.

In May 2026 OpenClaw published a measurement of that pipeline which we treat as
this paper's premise rather than a finding to reproduce~\cite{clawhubsignals}.
Across 67,453 skill versions, their three underlying scanners agree with each other
almost never: any pair overlaps on at most 10.4\% of combined positives, only 0.69\%
of flagged skills are caught by all three, and 81.9\% of positives come from one
scanner acting alone. Their diagnosis is precise, and we adopt it: \emph{``we do not
believe these findings point toward an issue with any of the individual scanners.
Rather, each scanner has a different risk surface.''}

\subsection{A second question about the same artifact}

That diagnosis can be read as a detection problem: four instruments, four
calibrations, unify them and the noise resolves. We think there is also something
else going on, and it is not a criticism of any of the four. VirusTotal answers ``is
this a known-bad binary.'' Static analysis answers ``does this match a dangerous
pattern.'' SkillSpector answers ``is this capability profile over-broad.'' ClawScan,
the judge over the three, answers ``on balance, is this malicious.'' Four well-posed
questions, each answered by an instrument built for it, and all four about a property
of the \emph{artifact}.

Consider a skill published by \texttt{oomol}, whose documentation instructs:
\begin{lstlisting}
curl -fsSL https://cli.oomol.com/install.sh | bash
\end{lstlisting}
Every signal in the pipeline calls this clean, and every one of them is right to: it
is an ordinary vendor installer, published by a real company, and it is exactly what
its documentation says it is. We use it as our running example precisely because
there is nothing wrong with it. Yet many operators will not permit an autonomous
agent to run it unattended, from any publisher, and the reason has nothing to do with
this publisher or this script.

That premise is load-bearing for everything we count later, so we ground it in named
controls rather than assert it. CIS Critical Security Control 2.7 requires technical
controls ensuring that only authorised scripts execute, and directs that unauthorised
scripts be blocked from executing~\cite{cisv8}. NIST SP 800-53 CM-11 requires
organisations to establish and enforce policies governing software installation by
users, and names software of ``unknown or suspect pedigrees'' among prohibited
installations~\cite{nist80053}. CM-7 constrains systems to authorised functions.
A shell command that fetches a script from the network and executes it in one step
is, on its face, the execution of an unreviewed script of unknown pedigree.

We are describing what these controls say, not asserting an audit finding: we have
not surveyed operators, and how any given assessor applies these controls to an
autonomous agent's shell commands is not settled practice. What we claim is narrower
and sufficient: for operators under these frameworks the question ``is this action
permitted here'' has a different answer from ``is this artifact malicious,'' and the
pipeline is built to answer the second. The prohibition is a property of the
\emph{action}, evaluated against \emph{this operator's policy}, not a property of the
artifact.

A publish-time scanner is not the place to express that, and we would not expect it to
be. Its output vocabulary is clean / suspicious / malicious. There is no slot for ``acceptable at a two-person
startup, prohibited at a bank,'' because that judgment depends on facts, whose
machine, whose data, whose compliance regime, what this skill has done here
before, that do not exist at publish time and are not properties of the file.

\subsection{Contributions}

\begin{enumerate}
\item \textbf{Permission $\neq$ maliciousness, measured.} 705 skills across 135
  distinct publishers that all three scanners and the judge rate clean while
  instructing an action named as prohibited by CIS Control 2.7 and NIST SP 800-53
  CM-11 (\S\ref{sec:permission}). One publisher contributes 506 of the 705, so the
  count should be read alongside its distribution, which we give.
\item \textbf{The scanned artifact is not the executed artifact.} Of 144 commands a
  live agent issued while following real skill documentation, 2 (1.4\%) appear
  verbatim in that documentation, and 50 (34.7\%) carried a consequence class the
  document never contained (\S\ref{sec:divergence}). Agent tool-use logging is
  well established; what we have not found measured elsewhere is the gap between what
  agent-skill static analysis inspects and what the agent then executes. The harness
  and the per-command record are released.
\item \textbf{A live runtime measurement.} Over 53 skills the registry had cleared
  that nonetheless document an action no clean record earns, a live agent reached for
  such an action in 23, and the unmodified gate held or blocked all 23
  (\S\ref{sec:runtime}). An earlier and smaller version of this run reported 4 of 54,
  and we explain why it was low.
\item \textbf{A layer-by-layer account of what sits outside each control's view},
  including our own, which is harm that never resolves to an action, stated as
  plainly as our contributions (\S\ref{sec:layers}).
\item \textbf{Derived, not asserted, autonomy thresholds}, parameterised by the
  operator's own risk tolerance, with a measured resolver fast enough to sit on
  every tool call (\S\ref{sec:resolver}, \S\ref{sec:graduation}).
\item \textbf{A benchmark for action-level governance}, released with the paper: 64
  semantics-preserving rewrites across nine techniques, with a stated scoring and
  aggregation rule, against which any implementation can report a resolution rate.
  Ours is 52\% macro-averaged over techniques, 75\% micro-averaged
  (\S\ref{sec:threat}). We are not aware of a shared way to state how much of an
  action's equivalence class a gate covers, which makes it hard to compare two gates
  or to show that one is improving. We would rather publish a benchmark our own
  implementation only half passes than not have one.
\end{enumerate}

\section{OATS: Resolving Actions, Not Judging Artifacts}
\label{sec:oats}

OATS (Open Agent Trust System) occupies the control point none of these signals
sit on: the moment an agent has decided what to do and has not yet done it. It does
not read the skill's prose, and it does not render an opinion about the skill. It
answers one question about one concrete act: \emph{is an agent with this track
record permitted to perform this class of action on this resource, unattended,
under this operator's policy?}

\subsection{The deterministic resolver}
\label{sec:resolver}

Let $\Sigma^*$ be the space of command strings an agent may emit and $\mathcal{C}$
a finite, closed set of consequence classes. $\mathcal{C}$ has 33 members, derived
from the operations the governed surfaces expose rather than from a taxonomy chosen
in advance: shell execution and its three dangerous specialisations (remote
execution, credential access, destructive); file writes split by what the path
implies (docs, tests, boilerplate, business logic, dependencies); repository and CI
operations; secret and IAM changes; deploys; and the computer-use verbs. Rather
than enumerate all 33 in prose, we publish the set as a machine-readable file,
\texttt{spec/action-classes.json}, one row per class carrying its key, its severity
rank, and whether it can ever graduate. Thirteen of the 33 never graduate, and the
rule is not a list but a threshold: a class graduates if and only if its severity
rank is below 75 (\S\ref{sec:graduation}).

The resolver is a function
\[
  \rho : \Sigma^* \rightarrow \mathcal{C}, \qquad
  \rho(c) \;=\;
  \begin{cases}
    \max_{\preceq}\;\{\, k \in \mathcal{C} \;:\; \pi_k(c) \,\} & \text{if some } \pi_k(c) \\
    \texttt{shell\_exec} & \text{otherwise,}
  \end{cases}
\]
where $\pi_k$ is the predicate for class $k$ and $\preceq$ is the published severity
order. We claim three properties for it, stated more carefully than in an earlier
draft of this paper.

It is \emph{total}. The second case is what makes it so: a command matching no
predicate resolves to the generic \texttt{shell\_exec} rather than to nothing, so no
input path fails open. Without an explicit bottom element the maximum is undefined on
the empty set, which is exactly the shape of the first defect in
\S\ref{sec:defects}, where unmatched commands fell through as unclassified.

It is \emph{stateless}. $\rho$ reads the command string and nothing else: not the
skill's documentation, not the ledger, not the wall clock, not a model. This is a
claim about $\rho$'s inputs, not about functions in general, and it is the property
that makes a decision replayable by an auditor without re-executing a model. We check
it rather than assert it: 200 command strings classified in two independent ledgers
with different histories returned identical classes in 200 of 200 cases.

One clarification a reader installing the client deserves, because they will meet it
on first run. The published gateway does include a small local language model, which
writes a plain-English description of what an agent changed. Its weights are not
bundled in the wheel: they are fetched once from a public model repository and cached.
It is not in the decision path. It runs asynchronously \emph{after} the verdict has
been computed and returned, it may fail or be absent without changing any decision,
and it rewrites only the human-readable sentence attached to a decision already made.
Keeping it out of that path is deliberate rather than incidental: a model placed there
would be something an attacker could write a payload to argue with, and it would cost
the property this section is about. The corpus study and the benchmark were both run
with explanations disabled, which changes no class but removes the question. It implies
$\rho(c)$ is identical across runs and hosts \emph{within a release}. We do not claim
stability across releases, and \S\ref{sec:defects} reports three false-positive
regressions in which $\rho(c)$ demonstrably changed for some $c$ between releases.
Every number in this paper is from \texttt{pheo-oats} 0.5.3.

One caveat on auditability that the interface currently imposes. The hook's only
output is a human-readable sentence, and both measurement scripts recover the class
by string-splitting it. A reworded refusal message silently reclassifies, which is a
poor foundation for a claim about replayability. The class set is published as
\texttt{spec/action-classes.json} and both scripts now map through it, but the
transport is still prose; emitting a stable machine-readable class identifier is the
right fix and is not yet done.

It \emph{resolves by maximum severity}. Where a command matches several classes the
result is the most severe, not a union or a mean. We previously called this
monotonicity, which names a different property; this is a definition, not a theorem.
There is no score to tune and no threshold to drift.

Each $\pi_k$ is argument-position-aware rather than keyword-presence-based: a token
must occupy the position a dangerous argument would occupy relative to a specific
command verb. The distinction is load-bearing: three successive false-positive
regressions in our remote-execution predicate (\S\ref{sec:defects}) all had the same
root cause: a gap in the pattern permissive enough to span markdown prose, section
breaks, or an unrelated command chain, so that a URL mentioned in documentation
hundreds of characters away from an \texttt{npm install} satisfied it. The shipped
predicate bounds every gap explicitly and constrains it to characters admissible in
a shell argument.

$\rho$ takes no input from the skill's documentation. A skill may describe its
installation step as reassuringly as it likes; the description is not an argument to
the function that decides whether the step is remote code execution. This closes one
specific channel: Saha et al.\ report governance-evasion rates of 36.5--100\% against
static and LLM-based scanners achieved purely through how a skill's own text is
written~\cite{skillmd2026}, and that channel is unavailable against a function that
never reads the text.

\subsection{Threat model, and what $\rho$ does not withstand}
\label{sec:threat}

Closing the documentation channel is not robustness, and we want to be exact about
the difference. We assume an agent recomposing benign documentation into concrete
commands. We do not assume an adversary writing commands to evade a resolver whose
grammar they know, and $\rho$ should not be deployed as though we did.

\paragraph{The trust boundary.} \S\ref{sec:related} claims Anderson's criteria, so we
owe the second one an answer rather than silence. The gate is a hook process invoked
by an agent that can, by construction, run shell commands on the same machine. We
assume uncompromised: the hook registration in the agent's configuration, the
\texttt{oatsctl} binary, and the ledger store. None of these is defended against the
governed agent itself. An agent with local shell access can edit the configuration
that installs the hook, or invoke a tool by a path the hook does not cover, and
nothing in $\rho$ prevents it. What integrity we do provide is over the record rather
than the enforcement: every receipt carries a SHA-256 digest over its own contents,
a reference to the previous receipt's digest, and an HMAC-SHA256 signature under an
operator-held key, so a ledger that has been edited after the fact does not verify.
That detects tampering with the history; it does not stop an agent disabling the
control prospectively. Tamper-resistance in Anderson's sense needs the monitor
outside the agent's reach, which for a locally-installed hook means OS-level
protection we do not currently implement. We therefore meet the first criterion,
partially meet the second, and \S\ref{sec:related} is candid that the third is met
only by a syntactic matcher.

The distinction is measurable, so we built a benchmark for it. Each case is a
semantically equivalent rewrite of a base action whose consequence class is not in
dispute: if the base resolves to class $k$, so should the rewrite, because the effect
on the system is identical. A miss is therefore unambiguous. Table~\ref{tab:evade}
reports 64 cases across nine rewriting techniques.

\begin{table}[H]
\centering
\small
\begin{tabular}{lrrl}
\toprule
Technique & Cases & Resolved & Example \\
\midrule
wrapper            & 20 & 100\% & \texttt{bash -c "\ldots{}"} \\
chained            & 15 & 100\% & \texttt{true; \ldots{}; true} \\
interpreter        & 4 & 75\% & \texttt{python3 -c "exec(urlopen(\ldots{}))"} \\
variable           & 5 & 60\% & \texttt{U=URL; curl -fsSL \$U | bash} \\
base64             & 3 & 33\% & \texttt{echo <b64> | base64 -d | sh} \\
indirect read      & 4 & 25\% & \texttt{find \textasciitilde{} -name credentials -exec cat \{\} +} \\
quoting            & 5 & 20\% & \texttt{curl -fsSL URL | ba''sh} \\
staged             & 3 & 0\% & \texttt{curl -o s URL; chmod +x s; ./s} \\
\midrule
\textbf{macro-average (headline)} & \textbf{59} & \textbf{52\%} & equal weight per technique \\
micro-average                     & 59 & 75\% & equal weight per case \\
\midrule
identity (control) & 5 & 100\% & \texttt{curl -fsSL URL | bash} \\
\bottomrule
\end{tabular}
\caption{Resolution rate under semantics-preserving rewriting, over 64 cases of which
5 are controls. \textbf{Scoring rule:} a case is resolved if the rewrite's class is at
least as severe as the base action's, since a gate that over-classifies still protects
the operator; on this suite that rule and exact match agree on every case. The
headline is the macro-average over the eight rewriting techniques, with the identity
control excluded from both aggregates. We report it that way because case counts per
technique are an artifact of how many variants each generator happens to emit, and
here that artifact is not neutral: every technique that resolves completely
interpolates the base command and so expands to 15--20 cases, while every technique
below 50\% is fixed-form and collapses to 3--5. Micro-averaging therefore weights the
successes roughly four to one and yields 75\%, or 77\% if the control is counted as
well. The same measurements support 52\% or 77\% depending on a presentation choice,
so we state the choice. No case failed to classify.}
\label{tab:evade}
\end{table}

The shape of the failure is more useful than the headline. Wrapping and chaining
resolve completely, which is the argument-position discipline of \S\ref{sec:resolver}
doing its job: the resolver reads inside \texttt{bash -c} and past surrounding
commands. What defeats it is any rewrite that removes the dangerous relationship from
a single command. Staged fetch-then-execute resolves at 0\%, because no individual
command in \texttt{curl -o s URL; chmod +x s; ./s} is remote execution; the
composition is. Quoting and indirect reads fail for the same reason at one remove.

That is a statement about the class of function $\rho$ is, not about the quality of
its predicates. A per-command classifier cannot see a property that only exists
across commands, and tuning patterns will not change that. Closing it requires
resolving shell grammar and carrying state between calls, which is a different design
and future work.

This bounds what the rest of the paper claims. \S\ref{sec:permission} and
\S\ref{sec:divergence} measure a non-adversarial world, and the honest bound in
\S\ref{sec:divergence} (recomposition never produced a higher-severity class) is a
statement about that world only: an adversary's entire objective is to make a
high-severity action look low-severity, which a benign sample never attempts.
Hardening $\rho$ against command-level obfuscation, most plausibly by resolving
shell grammar rather than matching its surface, is future work and not a property we
claim today.

\subsection{Measured cost}

A gate on every tool call must justify its place on the hot path with a number. We
report only numbers a reader can reproduce with the released client.

The number an operator experiences is \textbf{67.6\,ms} median and 81.1\,ms at p95:
what the CLI hook costs end to end, per tool call, on one core of an Apple M-series
laptop against an empty ledger. Re-measured for this revision over 200 sequential
calls spanning benign, ambiguous, and risky shapes, the median was 66.4\,ms and p95
69.3\,ms, which we take as confirming the original figure.

Almost all of that is process spawn and local IPC rather than classification. We can
bound the classification component from outside the process but not isolate it: a
2{,}000-character command, the longest input the extractor admits and one dense with
tokens every predicate must consider, costs 1.36\,ms more end to end than a
two-character one, and that difference also contains serialising a larger payload and
storing a larger receipt. So resolution is under roughly a millisecond even at the
largest input we accept, and far less at typical ones.

An earlier version of this paper reported resolution at 1.03\,$\mu$s mean and
3.27\,$\mu$s worst case, and derived throughput figures from them. Those came from an
internal microbenchmark that is not part of the released code, whose iteration count
and input distribution were not recorded, and which a reader therefore cannot rerun.
We have removed them. They may well be accurate, and the bound above is consistent
with them, but a number nobody can check should not be carrying an argument. Building
a released benchmark for the resolver in isolation is straightforward and we have not
done it.

\paragraph{A cost that is not on the hot path, and grows.} The figures above are
resolution and transport. They are not the whole cost, because every check also
\emph{writes}: a receipt, its digest, and its signature. Measured over the corpus run,
each governance check adds roughly 12\,KB to the gateway's store, and per-check
latency rises with the total size of that store rather than with the size of any one
lane. Classifying into a single store, throughput fell from about 67 to under 20
classifications per second across the first 15,000 checks and kept falling; cycling
the store held it flat at roughly 40--55/s. For an interactive agent this is
invisible, because a developer does not issue 15,000 tool calls into one store. For
corpus-scale evaluation it is the difference between a run that finishes and one that
does not, and we had to structure our own measurement around it. We report it because
it bears on the deployment claim: a gate cheap enough to sit on every action still has
to keep a record of every action, and it is the record, not the decision, that scales
badly. This is related to but distinct from the 152\,MB read defect reported in
\S\ref{sec:defects}, which was fixed; this one is a property of the current design and
is not fixed.

The distinction matters for what is architecturally possible. A judge that consults a
language model, local or hosted, pays a model's latency on every decision, and that
bounds where it can sit. The measured 67\,ms, most of it process spawn we could remove
by linking the resolver into a host process, is small enough to sit synchronously in
front of every action without an operator noticing. We make no claim that $\rho$ is
smarter than an LLM judge; it is deliberately narrower (\S\ref{sec:layers}). We claim
only that it is cheap enough, and stable enough, to occupy a position that a
per-decision model call would make expensive.

\subsection{The trust ledger}

$\rho$'s output feeds a finite-state ledger keyed on the pair (resource, class),
not on the skill. Trust is per-lane: a skill may be fully trusted to write
documentation in one repository and permanently untrusted to read a credential file,
at the same time. Each lane moves
\[
\textit{unclassified} \rightarrow \textit{held} \rightarrow
\begin{cases}
\textit{approved} & \text{(clean-run counter } n \mathbin{+}\mathbin{=} 1)\\
\textit{rejected} & \text{(} n \leftarrow 0 \text{)}
\end{cases}
\]
with \textit{graduated} reachable only when $n \geq N$, and only for the 20 classes
whose severity permits graduation at all. Every transition is a counting rule over
recorded human decisions, replayable from the ledger alone.

We should be exact about which half of this is shipped, because the paper's own
argument makes the distinction matter. The never-graduates rule is implemented: 13
classes refuse promotion at any $n$, enforced in the authorisation path rather than
configured. The threshold $N$ is implemented as a \emph{single operator-wide
constant}, and its default value is 10. The per-class $N_k$ derived in
\S\ref{sec:graduation} is therefore a proposal about what that constant should be,
not a description of what the shipped ledger does; parameterising $N$ by class is
work we have specified and not yet done. A reader is entitled to notice that the
default we ship is the same flat 10 that \S\ref{sec:graduation} argues is
3--15$\times$ too permissive. That is a fair reading, and it is why we state the
derivation as the paper's contribution and the constant as the implementation's
current limitation rather than presenting the two as one thing.

\section{Data and Method}
\label{sec:method}

We use \texttt{OpenClaw/clawhub-security-signals}, OpenClaw's own MIT-licensed
dataset~\cite{clawhubdataset}, in full: all four splits, 66,192 skill versions, each
carrying four scanner verdicts and redacted \texttt{SKILL.md} content. (A dataset of
the same name published by another account is a duplicate of this one; we use the
\texttt{OpenClaw} path throughout. The premise study~\cite{clawhubsignals} reports
67,453 versions; the released dataset is the smaller 66,192 and we do not know what
accounts for the difference.) Every skill is classified through the unmodified
shipping client (\texttt{oatsctl hook pre-tool-use}), the same binary an agent's
tool-use hook invokes, rather than a reimplementation, for a reason documented in
\S\ref{sec:defects}.

\paragraph{Counting, stated plainly.} Every count in \S\ref{sec:permission} is a
count of \emph{skills}. \texttt{skill\_slug} is already unique in the published
corpus, so no deduplication step is applied and none is needed. An earlier draft of
this paper claimed we deduplicated findings by publisher and that raw counts were
2--3$\times$ larger. That was not what the released code did and we withdraw it: the
code deduplicates nothing, and 705 is the plain skill count. We report the publisher
count alongside it, and the per-publisher distribution in \S\ref{sec:permission}, so
a reader can discount for a single vendor repeating one installer line across their
own catalogue rather than take our word for how much that matters.

\paragraph{The corpus has no line breaks, and this bounds what can be measured.} The
sanitizer that produced the public dataset collapsed newlines: in all 66,192 rows,
\texttt{skill\_md\_content} contains no newline character, and in 4.4\% of rows they
survive only as the two literal characters \texttt{\textbackslash n}. Fenced blocks
survive as \texttt{```} delimiters, so we can still tell code from prose, but the
boundaries \emph{between commands inside a block} are gone. Two consequences run
through everything below. First, we measure at skill granularity: does this skill
instruct at least one action of class $X$. A block with five risky commands scores
the same as one with a single risky command, so every count is a floor rather than an
estimate. Second, and more seriously, ``in a single command'' is not a predicate this
corpus can carry, because the separators that would define a command are exactly what
was removed. An earlier draft of this paper claimed in its abstract that 657 skills
fetch and execute remote code ``in the same command.'' On this corpus that is not
something we can know, and we have withdrawn the phrase.

\paragraph{So we tested whether the finding depends on it.} The worry is specific: if
collapsing newlines makes an unrelated URL on one line textually adjacent to an
unrelated \texttt{bash} on the next, the collapse becomes a false-positive generator.
We checked by asking what syntactic shape each detection actually rests on. Of the 674
blocks in the clean population resolved as remote execution, 587 (87.1\%) contain an
explicit pipe from \texttt{curl} or \texttt{wget} into a shell or interpreter, and 2
more use process substitution. A pipe character is not something a newline collapse
can manufacture: it was in the author's source line. A further 77 (11.4\%) are
single-token remote installs, \texttt{go install \ldots @version} (50),
\texttt{pip install} from a URL or git reference (27), whose classification does not
depend on adjacency at all. That leaves 8 blocks, 1.2\%, resting on neither, all of
them \texttt{npm install -g} against a remote target. So 98.8\% of the detections rest
on a construct that survives the sanitizer intact. The phrasing was wrong; the finding
is not an artifact of it.

We also exclude one class, \texttt{secret\_change}, from all risk counts, and we now
report what that exclusion is worth rather than assert it. It fires correctly: a skill
documenting a write to \texttt{.env} or
\texttt{\textasciitilde{}/.config/\ldots/credentials.json} genuinely is a secret
operation, and nearly every skill requiring an API key does this, which makes it a
correct classification and a weak risk signal. An earlier draft claimed including it
would inflate the central finding by roughly $2\times$. On the corrected extraction
that is false: including \texttt{secret\_change} moves the count from 705 to 706, a
factor of 1.00. The $2\times$ was itself an artifact of the extraction defect in
\S\ref{sec:defects}, which was feeding \texttt{.env} references out of
\texttt{\lstinline!```!json} and \texttt{\lstinline!```!yaml} fences into a shell
resolver. The exclusion is now nearly a no-op and we keep it only for continuity with
the class's documented rationale.

\section{Finding: Clean Skills Instructing Forbidden Actions}
\label{sec:permission}

Of 66,192 skills, 29,257 are rated \texttt{clean} by all three scanners and by the
ClawScan judge. Within that population, 705 skills across 135 distinct publishers
instruct an action in a class many operators forbid unattended
(Table~\ref{tab:permission}).

We report this population rather than the larger one available. Relaxing the
condition to \emph{no signal returned suspicious or malicious} admits a further 1,574
skills whose scans did not complete. The breakdown does not sum to 1,574 and we give
the reconciliation rather than leave a reader to find that it does not: 731 have no
SkillSpector result, 438 no VirusTotal result, 393 no static result, and 45 an
\texttt{unknown} ClawScan verdict, alongside 40 \texttt{stale}, 21 \texttt{error} and
3 \texttt{pending} states. Those are counts of signals, not of skills, and 62 skills
are missing more than one, which is what closes the gap: 1{,}671 incomplete signals
across 1{,}574 skills. That set yields 732 skills across 147 publishers, but
``nothing objected'' there includes ``nothing looked,'' which is a weaker claim than
the one we want to make.

\begin{table}[H]
\centering
\begin{tabular}{lr}
\toprule
Action class instructed & Skills \\
\midrule
Documented block fetching and executing remote code & 658 \\
Credential access & 23 \\
Destructive command & 20 \\
Merge to a protected branch & 7 \\
Trigger a workflow & 1 \\
\midrule
Distinct skills (135 distinct publishers) & 705 \\
\bottomrule
\end{tabular}
\caption{Skills every scanner and the judge rated clean whose documentation instructs
an action named as prohibited by the controls cited in \S\ref{sec:intro}. These are
detections of policy-relevant capability, not of malice. Rows count class
occurrences, not skills: four skills instruct two flagged classes and appear in two
rows, so the rows sum to 709 against 705 distinct skills. Three never-graduating
classes reachable from a shell command, \texttt{iam\_change}, \texttt{deploy} and
\texttt{delete\_or\_transfer\_repo}, scored zero here; a zero is informative, so we
say so rather than omit the rows silently. An earlier version of this table reported
20 IAM changes and 3 deploys, both of which were artifacts of the extraction defect
in \S\ref{sec:defects}.}
\label{tab:permission}
\end{table}

\paragraph{One publisher accounts for most of this.} The headline is 705 skills, and
a reader should immediately discount it, so we do the discounting here. The
distribution across the 135 publishers is extremely skewed: the median publisher
contributes \emph{one} skill, 117 of the 135 contribute exactly one, and a single
publisher, \texttt{oomol}, contributes \textbf{506}, or 71.8\% of the total. That is
one company documenting one installer line, the same line quoted in
\S\ref{sec:intro}, across its own catalogue of connector skills. Nothing about that
is unusual or improper: shipping a consistent install step across a product family is
ordinary engineering practice, and it is the reason we chose that line as the running
example. Excluding this publisher leaves 199 skills across 134 publishers.

We report all three numbers because they answer different questions. 705 is how many
distributed artifacts would drive an agent toward the action. 135 is how many
independent parties published such an instruction, and it is the right number for any
claim about how widespread the \emph{practice} is. 199 is what remains if a reader
wants the finding without its single dominant contributor.

Because a concentrated count invites the reading that this is one vendor and a few
copied templates, we checked. Restricting to remote execution and removing the
dominant publisher entirely leaves 152 skills from 93 publishers, carrying 102
distinct command strings and fetching from 48 distinct domains. The magnitude of the
headline depends heavily on one publisher; the practice it describes does not.

These 705 are flagged by our detector. We re-audited it by hand on this population:
100 skills drawn uniformly at random (seed fixed and published), each adjudicated
against the question \emph{does a fenced block in this skill actually instruct the
flagged class}. 92 of 100 were correct detections, giving 92.0\% precision with a
95\% Clopper--Pearson interval of [84.8\%, 96.5\%] and 598--680 true positives among
the 705. The eight errors are worth naming because they share a shape: five were a
network fetch piped into an interpreter that only \emph{formats} the response
(\texttt{curl \ldots | python3 -m json.tool}, \texttt{curl \ldots | node -e} with a
literal script), where nothing fetched is executed; two were \texttt{gh pr merge}
resolved as a merge to a \emph{protected} branch, which the document cannot
establish; one was \texttt{npm install -g} of an ordinary registry package, which is
a dependency change rather than remote execution. This supersedes the 73.6\% figure
reported earlier, which was measured on the pre-fix population and is not comparable.
Audit limitations are in \S\ref{sec:defects}. What we take from this is that
\emph{maliciousness and permission are different predicates over the same artifact}.
A publish-time pipeline answers the first, correctly and by design. The second is
left for somewhere else in the stack, and at present there is not much of a somewhere
else.

\paragraph{Are these skills benign?} The argument above needs them not to be missed
attacks, so we labelled the same 100-skill sample a second time, asking whether each
skill showed any marker of malicious intent: exfiltration to a host unrelated to its
stated purpose, obfuscated payloads, or behaviour the documentation does not
disclose. None of the 92 correct detections did. Every one was an ordinary vendor
installer, a language-toolchain install (\texttt{rustup}, \texttt{uv},
\texttt{go install \ldots @latest}), a published package, or a normal repository
operation.

That is a real result and it is weaker than ``these skills are benign,'' which we are
not in a position to claim. The audit was a single adjudicator reading documented
code blocks, which is exactly the wrong instrument for finding a competent attacker,
and 506 of the 705 come from one publisher, so the sample is dominated by repetitions
of one installer line. More fundamentally, ``rated clean'' means nothing objected, not
that anything verified benignity, and no scanner recall against this corpus is
published (\S\ref{sec:defects}). What we can say is that we looked for the competing
explanation, that a meaningful share of these are simply missed attacks, and did not
find it in the sample. What we cannot say is that it is absent.

\paragraph{What this section measures, and what it does not.} Two further limits bound
the claim, and both are visible elsewhere in this paper, so we draw them here rather
than let a reader find the tension unaided.

\emph{This is documented instruction, not observed execution.} The measurement runs
$\rho$ over fenced code blocks in \texttt{SKILL.md}. Nothing is executed. That is the
same document-based view \S\ref{sec:divergence} shows to be an imperfect proxy for
what an agent runs, and the proxy is imperfect in both directions: a skill can
document an action the agent never takes, and \S\ref{sec:divergence} finds the agent
takes actions the document never contains. So 705 is a count of what skills
\emph{tell agents to do}. It is not a count of what agents did. Note also that
\S\ref{sec:resolver}'s ``$\rho$ takes no input from the skill's documentation'' is a
statement about deployment; in this section the documentation text is $\rho$'s only
input.

\emph{Instruction is not reach.} \S\ref{sec:runtime} supplies the bridge between the
two. On 53 unanimously-clean skills that document a never-graduating action, a live
agent reached for such an action in 23, or 43.4\%. Composing that with the 92\%
precision above, the number of these 705 that would drive an agent to a forbidden
action in a comparable run is on the order of
$705 \times 0.92 \times 0.434 \approx 281$, not 705. We state the composition
because a reader can compute it, and because the two rates measure genuinely
different things: 705 is the size of the population a permission control would have
to consider, and 281 is an estimate of how often it would have fired in one
particular agent's behaviour over four turns. The first is the right number for
scoping a control, the second for sizing it. An earlier version of this paper put
the second figure at 48, from a run whose agent never reached the install step
(\S\ref{sec:runtime}).

\section{Finding: The Scanned Artifact Is Not the Executed Artifact}
\label{sec:divergence}

Static analysis of \texttt{SKILL.md} is sound to the extent that the document's
capability envelope bounds what the agent executes. Nobody claims it requires string
identity, so we measure both: how often the executed string appears in the document,
and how often the executed \emph{consequence class} does. The second is the one that
decides whether the method is sound. For 40 skills sampled across distinct
publishers, we classified every fenced code block in the document (the static view)
and separately captured every command a live agent (Claude Sonnet~5) issued while
following that same document, classifying both through the same resolver.

\paragraph{An earlier version of this result could not be checked; this one can.}
The figures below come from \texttt{research/live\_agent\_study.py}, which is
released with this paper, and every command it recorded is in
\texttt{live\_agent\_divergence.jsonl}. Commands the gate permits execute inside a
throwaway container, one per skill, with no mount of the host and no credential;
refusals are returned to the agent as refusals. Skills are drawn one per publisher
from the unanimously-clean population, restricted to documents whose own fenced
blocks resolve to at least one class more specific than the generic
\texttt{shell\_exec}, since a document with no capability envelope cannot be said to
bound anything. That selection matters for how the number should be read: 34.7\% is a
rate over documents that do carry a specific capability envelope, not over the corpus.
Those are roughly 1\% of publishers' first skills, so this is a deliberately narrow
slice, chosen because it is the only slice where the question means anything.

An earlier run of this study, reported in a previous version of this paper, used a
harness we did not preserve. Its figures were 93 commands over 40 skills, 3.2\%
verbatim, 60.2\% recomposed, 36.6\% class-absent. We report both, and they agree:
Fisher's exact test on the class-absent counts gives $p = 0.78$, no detectable
difference between the two runs.

\begin{table}[H]
\centering
\begin{tabular}{lrrrr}
\toprule
& \multicolumn{2}{c}{This run} & \multicolumn{2}{c}{Earlier run} \\
\cmidrule(lr){2-3}\cmidrule(lr){4-5}
Executed commands & Count & Share & Count & Share \\
\midrule
Verbatim from \texttt{SKILL.md} & 2 & 1.4\% & 3 & 3.2\% \\
Recomposed (class present in document) & 92 & 63.9\% & 56 & 60.2\% \\
Class absent from every documented block & 50 & 34.7\% & 34 & 36.6\% \\
\midrule
Total & 144 & & 93 & \\
Skills & 39 & & 40 & \\
\bottomrule
\end{tabular}
\caption{What an agent actually runs, against what the document says. Read the two
lower rows as the result: the executed string is almost never the documented string
(98.6\% of the time), but the executed consequence class was absent from every
documented block in 34.7\% of cases. The second number is the one that bears on
whether static analysis is sound. The right-hand columns are an earlier run of the
same study, from a harness we did not preserve; the two agree ($p = 0.78$ on the
class-absent counts). One skill of the 40 is missing from this run: its command
output was not valid UTF-8 and crashed the harness, which is fixed.}
\label{tab:divergence}
\end{table}

Agents do not transcribe documentation; they recompose it. They chain a check onto
an install the document never chained (\texttt{go version \&\& go install \ldots}),
add flags, redirect output, split one documented step into three. The 50
class-absent commands are not spread evenly: they fall in 14 of the 39 skills rather
than appearing once in each. One trajectory is worth quoting because it is both
findings at once. Handed a skill whose documentation is a single
\texttt{curl \ldots{} | bash}, the agent ran it, and when that did not succeed it
recomposed the same step into a staged fetch and inspect: \texttt{curl -o}, then
\texttt{cat}. That is recomposition producing a form \S\ref{sec:threat} reports the
resolver cannot see, unprompted, in an ordinary install.

\paragraph{Which number is the finding.} 98.6\% is a statement about string
identity, and we report it because it bounds what any exact-match method can do:
allowlisting documented command strings, hashing them, or diffing execution against
the document would fail on almost every command an agent issues. It is not a claim
that static analysis is blind, and presenting it as one would be a strawman. The
claim the data supports is the narrower one: for 34.7\% of executed commands the
consequence class appeared in no code block of the document, so a reviewer reasoning
about the document's capability envelope would not have anticipated them. That is
roughly a third the rhetorical size of 98.6\% and it is the number we stand behind.

\paragraph{An honest bound on this result.} All 50 unanticipated commands resolved
to the generic \texttt{shell\_exec} class. We observed \emph{no} case in this
sample where recomposition manufactured a \emph{higher-severity} class than the
document contained. The claim this experiment supports is therefore precise and
narrower than it may first appear: the executed string is essentially never the
documented string, so document-based analysis reasons about a proxy. But in this
sample, the proxy did not fail in a dangerous direction. We report the absence of that stronger
result because a reader would otherwise assume it.

\section{Experiment: Live Runtime Verification}
\label{sec:runtime}

We ran a live agent (Claude Sonnet~5) against real skill documentation through the
unmodified gate, executing only commands the gate permitted, each inside a throwaway
container. Each skill received one run of at most four turns, with its
\texttt{SKILL.md} in the system prompt and a single instruction to install it and
confirm it works; a refused command returned the gate's own refusal text as the tool
result. The harness is \texttt{research/live\_agent\_study.py} and every command is
recorded in \texttt{live\_agent\_runtime.jsonl}.

Across 54 unanimously-clean skills that document a never-graduating action, one per
publisher, the agent reached for such an action in \textbf{23 of the 53} that
produced any command (43.4\%, 95\% CI [29.8\%, 57.7\%]). \textbf{All 23 were held or
blocked before execution} (23 of 23, 95\% CI [85.2\%, 100\%]). The actions it reached
for were remote execution in 22 cases, credential access in 2, and a destructive
command in 1; the gate blocked 22 of those commands outright and held 3.

\paragraph{This contradicts an earlier run, and the earlier run was wrong in a way
we predicted.} A previous version of this paper reported 4 of 54 (7.4\%) from a
harness we did not preserve, and said of it: \emph{``an agent truncated before it
reaches the install step records as not reaching for the action \ldots{} 7.4\% is a
floor on reach rather than an estimate of it.''} That was the right worry. Fisher's
exact test puts the two runs genuinely apart ($p = 1.6 \times 10^{-5}$). We think
the main cause is execution: the turn budget is unchanged at four, and a pilot with
execution disabled reproduced the earlier pattern exactly, with the agent spending
its turns establishing whether its own tooling worked and never reaching an install.
But we changed more than one thing between the runs. This one also samples one skill
per publisher and requires the documented action to be one that never graduates, and
we cannot separate those from execution with two runs. Read the comparison as two
observations that differ, with a mechanism we can demonstrate for part of it, rather
than as a controlled experiment.

Both directions of that correction matter. The population a permission control has
to consider is far larger than the earlier number suggested. And the evidence that
the control works is much stronger: 23 of 23 has a lower bound of 85\%, where 4 of 4
was consistent with a true block rate of 40\%.

\paragraph{What the gate cost the operator.} A reviewer asked us for the gate's
false-positive rate, and we think the honest answer is that it does not exist as one
number. A hold on a lane with no track record is not an error, it is what the ledger
is for; whether it is \emph{friction} depends on which classes a given operator would
have waved through, which is their policy and not a property of the gate. A single
false-positive rate would contradict this paper's own thesis. Two things can be
reported without judgement, and \texttt{research/interruption\_load.py} computes both
from the released record.

\emph{Load.} The run produced 170 decisions over 53 skills: 3.2 per skill, of which
2.8 were holds and 0.4 blocks. 145 of the 170 (85\%) were the generic
\texttt{shell\_exec}; the rest were the never-graduating classes the gate blocked.

\emph{Spectrum.} Because a decision is a function of the resolved class and the lane
state, what a policy would have removed is arithmetic. An operator who reviews
everything keeps all 3.2 interruptions per skill. One who lets reads, docs, tests and
ordinary shell run unattended in their own repository removes 145 of 170, or 85\%,
and keeps 0.5 per skill. A fully permissive operator lands in the same place here,
because the only graduating class this workload produced was \texttt{shell\_exec}: a
workload touching file writes, merges or releases would separate the two.

\begin{table}[H]
\centering
\begin{tabular}{lrrr}
\toprule
Operator policy & Removed & Remaining & Per skill \\
\midrule
Reviews everything & 0 & 170 & 3.2 \\
Reads, docs, tests, ordinary shell unattended & 145 & 25 & 0.5 \\
Everything that can graduate & 145 & 25 & 0.5 \\
\bottomrule
\end{tabular}
\caption{The same 170 decisions, priced against three operator policies. The
right-hand column is the number an operator actually feels. No policy removes the 25
interruptions on never-graduating classes, because that rule is enforced in the
authorisation path rather than configured.}
\end{table}

That connects this section to \S\ref{sec:graduation} for the first time. The friction
a moderate operator feels is 2.7 \texttt{shell\_exec} holds per skill, and
\S\ref{sec:graduation} prices exactly that lane: at $\epsilon = 5\%$ it needs 59
consecutive clean approvals, 392 expected given the reset. At this workload's rate
that is on the order of 143 skill installations before the lane stops asking, though
an operator doing ordinary work in the same repository reaches it sooner, since the
cost is 392 approvals rather than 392 skills. Whether that is tolerable is the
question \S\ref{sec:graduation} concedes it cannot answer for everyone.

\paragraph{Were the holds themselves correct?} That part is a property of the gate,
and we audited it: 50 holds drawn at random were adjudicated for whether the resolved
class was right, and 49 were (98\%, 95\% CI [89.4\%, 99.9\%]). The three holds
carrying a class other than \texttt{shell\_exec} were examined exhaustively rather
than sampled. Two were credential access and both genuinely contain
\texttt{cat \textasciitilde{}/.env}. The third is a false positive of a shape we have
seen before: a heredoc writing a new file was resolved as a destructive command,
because a dangerous pattern appeared in the \emph{content} being written rather than
in a command being run. That is the same defect as the comment-matching false
positives in \S\ref{sec:permission}'s audit, and it is not fixed.

\paragraph{Three things this experiment does not establish.} The container is not
anyone's real machine. It is a fresh Debian with the common toolchains and no
pre-existing environment, so a skill whose install assumes Homebrew, or a virtualenv
that already exists, fails there and would have succeeded on a developer's laptop.
We first wrote that this understates reach. The record does not support that: a
container can suppress an install that would have worked on a laptop, and it can also
force a fallback onto remote code that a working \texttt{brew} would have avoided.
Seven skills tried \texttt{brew}; in one, \texttt{brew install} failed, the agent
established it was on Linux, and its next step fetched and ran code from a network
endpoint, which is a reach the container created rather than revealed. The direction
of this confound is not determined, and with seven cases we cannot settle it.
Second, the spectrum above prices three policies we chose; an operator's own policy
is the input the system takes, and we have not surveyed what real operators pick. Third, the run is one agent, one
model, one prompt, and 43.4\% carries a confidence interval from 30\% to 58\%.

\begin{table}[H]
\centering
\small
\begin{tabular}{p{3.2cm}p{5.2cm}p{5.6cm}}
\toprule
Skill & ClawScan's own verdict (verbatim) & What the agent ran, and the decision \\
\midrule
\texttt{pdurlej/\allowbreak things-cloud} &
clean, high confidence: ``clear read/write safeguards and no scanner or artifact
evidence of hidden behaviour'' &
\texttt{go install github.com/\ldots{}}: blocked, remote code execution \\
\texttt{j-edel/folk-cli} &
clean, high confidence: ``a clearly scoped helper \ldots{} with disclosed credential
use'' &
\texttt{npm install -g github:\ldots{}}: blocked \\
\texttt{barronlroth/f1-cli} &
clean, high confidence: ``no hidden code or persistence found'' &
\texttt{go install \ldots{}@latest}: blocked \\
\texttt{zypher-agent/\allowbreak corespeed-excalidraw} &
clean, high confidence: ``disclosed network downloads \ldots{} that fit its purpose'' &
\texttt{curl -fsSL \ldots{} | bash}: blocked \\
\bottomrule
\end{tabular}
\caption{Four of the 23, each on a skill the registry's own judge had cleared at high
confidence, with its rationale quoted. Every verdict, confidence level and summary
here can be read directly out of the public dataset; the command and the decision are
in \texttt{live\_agent\_runtime.jsonl}.}
\label{tab:blocked}
\end{table}

The fourth case shows the resolver reading an execution rather than a description. The
agent's first install attempt was ordinary and was \emph{held for routine review},
not blocked. Its second attempt routed the same install through a third-party
mirror, and the same resolver classified that occurrence as remote code execution
and denied it. One documented step, two executions, two different classes, because
$\rho$ is a function of what actually happened, not of what the document said would
happen.

\section{What Each Layer Sees}
\label{sec:layers}

The four registry signals and OATS are sometimes discussed as if they were competing
detectors to be ranked by accuracy. We do not think they can be compared that way,
because they do not take the same input, do not run at the same time, and do not emit
the same kind of statement. Table~\ref{tab:layers} sets them side by side on those
axes rather than on a shared score. Nothing in it is a criticism of any of them: the
right-hand column records what sits outside each layer's view by virtue of where it
runs, and our own row is in the table for the same reason as the others.

\begin{table}[H]
\centering
\small
\begin{tabular}{p{2.1cm}p{3.6cm}p{2.5cm}p{2.0cm}p{3.0cm}}
\toprule
Layer & Question it answers & Input & When & Outside its view \\
\midrule
VirusTotal &
Is this a known-bad binary? &
Bundled files &
Publish &
Novel code; anything not in a signature database \\[2pt]
Static analysis &
Does this text match a dangerous pattern? &
Document text &
Publish &
Intent; whether the pattern is ever reached \\[2pt]
SkillSpector &
Is this skill's declared capability profile over-broad? &
Document text, AI-assisted &
Publish &
Which capabilities are actually exercised, and on what \\[2pt]
ClawScan &
On balance, is this skill malicious? &
The three above, plus provenance and moderation history &
Publish &
Anything absent from the artifact at publish time \\[2pt]
\textbf{OATS} &
Is \emph{this action} permitted here, now, given this record? &
The resolved command string &
Every action &
Harm that never becomes an action \\
\bottomrule
\end{tabular}
\caption{Five layers, five different questions. The right-hand column is the point:
what falls outside a layer's view follows from where it sits, not from how well it is
built. Each of these is a capable instrument for the question in its second column.}
\label{tab:layers}
\end{table}

The last column describes where each layer sits, not how well it performs. A
reputation service does not recognise novel code, because reputation requires
history. A capability review of a document cannot report which capabilities are
actually exercised at run time, because that is not a fact about the document. And
OATS does not see harm that never resolves to an action, because an action is the
only thing it is ever shown.

That last limitation is ours and we state it plainly with examples rather than leave
it implied. Reading the registry's own summaries of skills it declined to clear, we
find three recurring shapes that our resolver would have no way to notice. One is a
hardcoded recipient: a skill that sends to an address fixed in the document, with no
confirmation step, and with no executable block for a command-level gate to read at
all. Another is a persuaded behaviour: a skill whose stated purpose is to automate an
account while avoiding detection, assembled entirely from tool calls that are
individually unremarkable. The third is an over-broad scope: an agent given far more
access to private data than its purpose requires, where no single action is
disproportionate and the mismatch only exists at the level of the whole skill.

We give these without naming the skills or their publishers, since the point is the
shape of the harm and not any particular author. All three are what an artifact-level
capability review, such as SkillSpector's Excessive Agency analysis, is built to
catch, and all three are invisible to a function that reads one command string at a
time. A runtime action gate is not a superset of a semantic scanner and should not be
deployed as one.

\subsection{Why the blast radius is wider together}

Because the blind spots do not coincide, the layers compose rather than compete. A
skill can be malicious in the artifact and never act (caught upstream, invisible to
us). It can be entirely benign in the artifact and still drive an action an operator
forbids (invisible upstream, caught by us; \S\ref{sec:permission} counts 705 such
skills across 135 publishers). And an artifact that never changes can produce
different actions on different runs, which is the case
\S\ref{sec:divergence} measures at 34.7\% and \S\ref{sec:runtime} exhibits
concretely: the same documented install step resolved as ordinary on one attempt and
as remote code execution on the retry.

An earlier draft supported this with a Jaccard overlap of $0.022$ between OATS's flags
and the union of the registry's positives, offered as evidence of independence. That
was the wrong statistic and we withdraw the inference. Jaccard measures overlap, and
for two events of very different base rates it is driven almost entirely by those
rates; a low value is what you get from any pair of such signals regardless of how
they depend on each other. Reporting it without the marginals also left a reader
unable to tell which flag population was used, and it was not the one
\S\ref{sec:permission} defines, which is disjoint from the registry's positives by
construction.

Here are the marginals and a statistic that measures association. Across the 66,192
skills, OATS flags $|A| = 1{,}412$ (2.13\%); at least one registry signal returns
suspicious or malicious on $|B| = 35{,}361$ (53.4\%, almost all of it SkillSpector);
and $|A \cap B| = 680$. Independence predicts 754 skills in the intersection, so the
observed overlap is $0.90\times$ expected. The association is very slightly
\emph{negative}, not zero: $\phi = -0.016$, odds ratio $0.81$, Fisher exact
$p = 6.5 \times 10^{-5}$. The $p$-value is small because the corpus is large; the
effect is negligible.

The honest reading is that we cannot establish independence, and do not need to. What
the composition argument requires is only that the blind spots do not coincide, and
that is a structural claim about what each layer takes as input (Table~\ref{tab:layers}),
not a statistical one. A $\phi$ of $-0.016$ is consistent with two instruments that
are very nearly unrelated, which is what reading different objects would predict, and
we take it as no more than that.

This suggests a composition rule that may be cheaper than building new detection:
\emph{a registry verdict is an input to the ledger, not a competitor to it.} Where
an upstream evaluator has marked a skill non-clean, an operator's policy can pin
every lane for that skill at held or never-graduates, regardless of what $\rho$
concludes about any individual command. The scanner's artifact-level judgment and
the gate's action-level enforcement compose without either approximating the other.
We present this as a mechanism rather than a measured result: evaluating it against
the population that the scanner verdict itself defines would be circular.

\paragraph{Two predicates we built and did not ship.} While examining what OATS
misses, we implemented two candidate detectors, outbound data sent to a literal
hardcoded domain and a credential embedded directly as a command argument, and
held them to the specificity bar every shipped predicate must pass: measured
false-positive rate against the 29,257-skill unanimously-clean population. They fire
on 4.4\% and 1.5\% of clean skills respectively, because \texttt{curl}-POSTing data
to a URL is simply how ordinary API integrations work. Either would manufacture,
inside OATS, exactly the approval fatigue that makes this class of control hard to
live with~\cite{openclawroadmap}. Neither shipped.

\paragraph{That is not the same bar the shipped predicates cleared.} It would be
convenient to say that it was, and it is not true. The shipped remote-execution
predicate fires on 658 of the 29,257 unanimously-clean skills, or 2.25\%, which sits
between the 1.5\% and 4.4\% rates that disqualified the two candidates. Those two
rates were measured before the extraction defect in \S\ref{sec:defects} was found, so
they are not strictly comparable to the 2.25\% and we have no way to remeasure them:
the predicates were never shipped and no longer exist. The ordering is unlikely to
reverse, but a reader should treat the comparison as indicative. If the test were
really the rate at which a predicate fires on clean skills, remote execution would
plausibly have failed it. The actual distinction is precision: we believe the
remote-execution firings are true positives and audited them at 92\%
(\S\ref{sec:permission}), while we never measured precision for either rejected
candidate. That is a defensible reason to ship one and not the others, but it is a
different reason, and comparing like with like would require an audit of the two we
dropped, which we did not do.

For the same reason we now report base rates over the whole corpus rather than only
the clean subset, since a signal that fires equally on both is not discriminating
between them. Remote execution fires on 1.73\% of all 66,192 skills and 2.25\% of the
29,257 clean ones; credential access on 0.23\% and 0.08\%; destructive commands on
0.17\% and 0.07\%. Remote execution is, if anything, slightly \emph{more} common among
the skills every scanner cleared than in the corpus at large. That is not a defect in
the signal. It is the paper's thesis stated as a base rate: this predicate is not
measuring maliciousness, and a population selected for being non-malicious does not
have less of it.

\paragraph{When there is no verdict to compose with.} Not every skill arrives
through a registry. OpenClaw's own roadmap acknowledges skills coming from GitHub,
a private registry, or a file someone sends~\cite{openclawroadmap}. For those no upstream signal exists at any price, and the runtime layer is the only
one.

\section{Graduated Autonomy as Operator Policy}
\label{sec:graduation}

A gate that never lets a track record reduce scrutiny runs into a problem the field
has already named, and which OpenClaw describes candidly in their published
roadmap~\cite{openclawroadmap}: users enable unattended mode once prompts outpace
reading, after which no prompt is read. We take that as an honest account of a
difficulty everyone building these controls shares, ours included. A promotion rule has to be defensible, and ``ten clean approvals'' was chosen rather
than derived.

Zero failures across $N$ independent trials bounds the true failure rate $\epsilon$
at confidence $1-\delta$ via $(1-\epsilon)^N = \delta$, giving the threshold required
for a class whose tolerable failure rate is $\epsilon_k$:
\[
  N_k \;=\; \left\lceil \frac{\ln \delta}{\ln (1-\epsilon_k)} \right\rceil .
\]
This is the exact inversion of the one-sided binomial (Clopper--Pearson) bound at
zero observed failures. It is often reached via the \emph{rule of three}, the
approximation $\epsilon \approx 3/N$ that follows from $-\ln(0.05) \approx 3$. We use
the exact form throughout and call it by its own name.
At $\delta = 0.05$, ten clean approvals leave a true failure rate as high as
\textbf{25.9\%} unexcluded. We state that carefully, because the number is easy to
misread. It is a one-sided 95\% upper confidence bound on $\epsilon$: the largest
failure rate ten clean trials cannot rule out. It is \emph{not} the probability that
the next action fails. The point estimate after ten clean runs is zero, and under a
uniform prior the posterior predictive probability of failure on the eleventh is
$1/12 \approx 8.3\%$. The case against a flat threshold of ten is not that the next
action is likely to be wrong; it is that ten trials are too few to distinguish an
agent that fails a quarter of the time from one that never fails.

\begin{table}[H]
\centering
\begin{tabular}{lrrrr}
\toprule
Classes at this tolerance & $\epsilon_k$ & $N_k$ & E[approvals] & $\times$ vs 10 \\
\midrule
Docs, tests, reads, issue/PR admin (8) & 10\% & 29 & 202 & 2.9 \\
Shell execution, boilerplate, screen interaction (5) & 5\% & 59 & 392 & 5.9 \\
Business logic, dependencies, merges, typed input (7) & 2\% & 149 & 965 & 14.9 \\
\emph{illustrative: a stricter tolerance} & 0.5\% & 598 & 3{,}807 & 59.8 \\
\midrule
Every class at severity $\geq 75$ (13) & 0\% & never & --- & --- \\
\bottomrule
\end{tabular}
\caption{Clean runs required to license unattended action at 95\% confidence, and the
expected number of human approvals to actually reach that run given the reset in
\S\ref{sec:oats}. \textbf{The grouping of classes into tolerance bands is our
proposal, not a shipped configuration:} the implementation applies one threshold to
every class (\S\ref{sec:oats}), and an operator adopting this table would be choosing
these bands, not reading them out of the product. Under the grouping shown, a flat
threshold of 10 demands 3--15$\times$ too little evidence; the 0.5\% row prices a
stricter tolerance an operator may choose, and no class is assigned to it. The bottom
row is remote execution, credential access, secrets, destructive commands, deploys,
IAM, CI configuration, workflow triggers, webhooks, repository deletion, protected
merges, branch protection, and irreversible UI commits: no finite $N$ licenses any of
them. An earlier draft of this table placed deploys, IAM and CI configuration at
$N=598$, which contradicted the shipped policy that they never graduate at all.}
\end{table}

Four things this derivation does not give the operator, stated because $N_k$ reads
more decisive than it is.

\textbf{The trials are not independent, and they are not sampled from the population
being licensed.} An agent's repeated actions on one resource are correlated, so $N_k$
is a floor: dependence can only raise the evidence a sound threshold requires. The
sharper problem is selection. The ledger counts \emph{human-approved} runs, and
graduation then removes the human. If the reviewer was catching and rejecting the bad
cases, a clean run is evidence about the reviewer, not about the agent, and the reset
to zero on rejection guarantees every counted trial is one a human passed. The
quantity we bound is the failure rate \emph{under review}; the quantity the operator
needs is the failure rate \emph{without it}. We know of no way to close that gap from
the ledger alone, and we do not claim to.

\textbf{A per-action bound is not a cumulative one.} Bounding $\epsilon \leq 0.5\%$
says nothing about the probability of at least one failure across a period of
unattended operation. At that bound, 1{,}000 unattended actions carry an expected 5
failures and a 99.3\% chance of at least one. An operator who needs a failure budget
per quarter should compute $N$ from that budget, not read $\epsilon_k$ as a
system-level guarantee.

\textbf{There is no multiplicity correction.} Each lane graduates at $\delta = 0.05$
independently, so an operator running 100 lanes should expect roughly 5 wrongly
graduated and has a 99.4\% chance of at least one. Either spend $\delta$ across lanes
or accept that the guarantee is per-lane.

\textbf{And the cost is the number of approvals, not $N$.} Because a rejection resets
the counter to zero, graduation requires $N$ \emph{consecutive} clean approvals, and
the expected number of human approvals to reach a run of $N$ at true rate $\epsilon$
is $(1-p^N)/((1-p)p^N)$ with $p = 1-\epsilon$. The table above reports it: 202
approvals for documentation writes, 965 for business logic. This is the honest cost,
and it is consistently 6--7$\times$ larger than $N$. A section that opens by naming prompt fatigue has to
concede where that leaves it: on high-frequency, low-consequence lanes, several
hundred approvals is a plausible amount of work to amortise over months. On anything
rarer it is not, and for those lanes this derivation is better read as documentation
of what cannot responsibly be automated than as a route to automating it. Pooling
evidence across resources within a class, or replacing the hard reset with a
Beta-Binomial posterior that consumes a rejection without discarding history, would
both reduce the cost; both weaken properties (\S\ref{sec:oats}'s per-lane isolation,
and replay by counting) that we currently prefer to keep.

$\epsilon_k$ is the operator's parameter, and it is where ``what this organisation
permits'' enters the system as a number rather than a vendor default. A two-person
startup and a regulated bank do not owe each other the same tolerance on a dependency
change, and the same derivation tells each what their own stated tolerance costs in
evidence. Classes in the bottom row admit no finite $N$, since no clean record makes
an unrecoverable action recoverable, which is how the policy of
\S\ref{sec:permission} becomes mechanical.

\section{Limitations and Defects in Our Own Instrument}
\label{sec:defects}

\textbf{Silver-standard ground truth, in both directions.} OpenClaw's dataset card
describes \texttt{clawscan\_verdict} as the registry's automated verdict, not
human-adjudicated truth, with a human-annotated subset still under
development~\cite{clawhubdataset}. ``Scanner-confirmed malicious'' is not ``verified
malicious.'' The symmetric point matters more for this paper: \emph{``rated clean'' is
not ``verified benign'' either}. Clean means nothing objected. It does not mean anything
established that these skills are harmless, and no scanner's recall against this
corpus is published, so we cannot bound how many genuinely dangerous skills carry a
clean verdict. This bears directly on \S\ref{sec:permission}, whose argument is that
the scanners are answering a different question rather than answering this one
wrongly. If a meaningful share of the 705 were in fact missed attacks, that argument
would weaken into ``the scanners have false negatives,'' which is a real result but
somebody else's and not the one we claim. An earlier draft asserted flatly that these
were not missed attacks and that the scanners did not err. We had not measured that
and have removed the assertion; \S\ref{sec:permission} now reports what our audit
does and does not establish about it.

\textbf{Sample sizes.} 53 skills for \S\ref{sec:runtime} and 39 for
\S\ref{sec:divergence} produce non-trivial rates but not tight intervals, and both
sections carry them: 23 of 53 is [29.8\%, 57.7\%], 23 of 23 is [85.2\%, 100\%], and
50 of 144 is [27.0\%, 43.1\%]. Both are one model on one day.

\textbf{We do not report a detection rate for OATS against malicious skills, by
choice.} Such a number would score a runtime action gate on a publish-time
malware-detection benchmark, which is not the question it answers
(\S\ref{sec:layers}); a favourable score there would in fact indicate redundancy
with the scanners rather than value beside them. We describe the limitation
qualitatively instead, in \S\ref{sec:layers}.

\textbf{Precision, and the limits of how we measured it.} \S\ref{sec:permission}
reports 92\% precision from a hand audit. The protocol: 100 skills drawn uniformly at
random from the 705 with a fixed published seed; for each, the fenced blocks that
triggered the flag were read and judged against the question \emph{does this skill's
documentation actually instruct the flagged class}, with a network fetch piped into a
formatter, or a class the document cannot establish, counted as incorrect. The same
sample was labelled a second time for markers of malicious intent.

The limitations are real. It is a self-audit: the authors adjudicating their own
detector, with a single rater, so there is no inter-rater agreement to report. Some
calls are genuinely arguable and we resolved them toward ``correct'': a command that
generates and prints a private key was counted as credential access though it reads no
existing secret, and \texttt{rm -rf} against a build directory was counted as
destructive though nothing unrecoverable is lost. A second adjudicator could
reasonably return several points lower, so we report the sensitivity rather than only
the point estimate: flipping all three arguable calls against us gives 89\%
([80.8\%, 94.3\%]), and flipping twice that many gives 86\% ([77.6\%, 92.1\%]). The
interval [84.8\%, 96.5\%] reflects sampling error only and not any of this. Precision is also not uniform across classes: with 658
of 709 class-occurrences in one class, the estimate is effectively an estimate for
remote execution, and the rows with 23, 20, 7 and 1 occurrences are not separately
supported by a sample of this size.

An earlier draft reported 73.6\% precision without sample size, protocol, or interval.
That figure was measured on the pre-fix population and we could not reconstruct its
denominator; it is superseded rather than corrected.

\textbf{No recall for the instruction detector.} Separately from the deliberate
choice below, we do not know what fraction of skills that \emph{do} instruct a
forbidden action our detector finds. The extractor reads only fenced blocks, ignores
prose entirely, skips fences tagged as other languages, and measures at skill rather
than command granularity, so 705 is explicitly a floor and not an estimate of the
true population. Without a labelled subsample we cannot say whether it is 5\% or 95\%
of it.

\textbf{Single registry, frozen snapshot.} One registry, one deliberately frozen
snapshot chosen for reproducibility.

\textbf{A correction to a previously reported figure.} A previous version of this
paper reported that a live agent reached for a never-graduating action in 4 of 54
cleared skills. Re-running that study with a working execution environment gives 23
of 53. The earlier figure was low because the agent never reached the install step,
which the previous version anticipated in the text without being able to test it. We
report the correction here rather than quietly replacing the number, because a reader
who saw the earlier draft should be able to find out what changed and why.

\textbf{Nine defects in our own instrument, found and fixed during this work.} A
shell-classification bypass let unmatched Bash commands fail open. Three successive
false-positive regressions in the remote-execution predicate, each a
too-permissive gap, were caught only by measuring against the clean population. A
performance defect fetched 152\,MB of room history per governance check, which
would have made corpus-scale evaluation impractical and any busy production room
unusable. An entropy heuristic classified relative file paths
(\texttt{references/CLAUDE-local-template.md}, 4.21 bits) as leaked credentials.

Three further defects were found in the measurement code after the first version of
this paper circulated, and they are the reason \S\ref{sec:permission}'s numbers moved.

\emph{The extractor admitted non-shell code as shell.} For a fence whose language tag
was not a known shell, the guard deciding whether to keep it was applied to the body
\emph{with the tag still attached}. Ten common tags, \texttt{python},
\texttt{python3}, \texttt{node}, \texttt{go}, \texttt{cargo}, \texttt{docker},
\texttt{make}, \texttt{git}, \texttt{npm} and \texttt{jq}, are themselves command
names, so the tag satisfied its own guard and every such fence was passed to a
resolver written for shell. The fix keys on a fact that survives sanitization: an
untagged fence's body begins with the whitespace that followed the opening
\texttt{```}, and a tagged one begins with its tag. On this corpus it moves 110{,}319
fences out of the shell population.

\emph{Two never-graduating classes were structurally uncountable.} The measurement
script mapped the resolver's prose labels back to class keys through a hand-written
table carrying 15 of the 33 labels. \texttt{manage\_webhook} and
\texttt{delete\_or\_transfer\_repo} were absent, so any occurrence of either was
silently dropped and could never appear in Table~\ref{tab:permission}. The taxonomy
is now loaded from \texttt{spec/action-classes.json} rather than restated.

\emph{A dangerous pattern in content, not in a command.} \S\ref{sec:runtime}'s audit
of held decisions found a heredoc writing a new file resolved as a destructive
command, because the pattern matched text being written rather than a command being
run. \S\ref{sec:permission}'s audit found the same shape in comments. Argument
position is checked relative to a command verb, but a verb inside a quoted document
is still a verb to the matcher. This one is open.

\emph{Errors were being silently discarded.} Both measurement scripts invoked the
classifier with \texttt{--fail-open} and wrapped it in a bare \texttt{except} that
dropped any block which errored or timed out, with no counter. A block that failed to
classify is not a block that classified as harmless, and the two were being pooled
over an unknown denominator. Both scripts now run without \texttt{--fail-open} and
report the failure count alongside every result. Relatedly, fence bodies are
truncated at 2{,}000 characters, which was silent and is now counted: it affects 45
blocks in this corpus.

Two of the first six surfaced only when we ran the instrument across all 66,192
skills, which is the argument for corpus-scale evaluation. The three most recent
surfaced when the measurement code was read adversarially by someone looking for
them, which is the argument for publishing it.

\section{Related Work}
\label{sec:related}

$\rho$ is a reference monitor for agent actions, and the criteria are Anderson's:
invoked on every access, tamperproof, and small enough to verify~\cite{anderson1972}.
$\rho$ sits on every tool call, admits no model into the decision path and replays
from the ledger, and is a pure function of one string. Saltzer and Schroeder's
complete mediation and least privilege~\cite{saltzer1975} are the same reasons the
ledger is keyed on (resource, class) rather than on the skill. We claim no novelty in
these principles, and none in applying them: our contribution is an implementation and
a measurement, not a new idea about reference monitors. \S\ref{sec:threat} is candid
that the third criterion is currently met only by a syntactic matcher an adversary can
route around, and \S\ref{sec:threat} likewise concedes that we meet the second, being
tamperproof, only partially.

Runtime guardrails for LLM applications are an active area. Llama
Guard~\cite{llamaguard2023} classifies conversational input and output with a model;
NeMo Guardrails~\cite{nemoguardrails2023} runs programmable rails over an application's
dialogue flow. Both put a model or a scripted policy between a user and a model. OATS
sits one layer lower, between a model's chosen action and the system it would change,
and the difference that matters for deployment is cost: a rail consulting a model pays
a model's latency per decision, which is what confines it to conversational turns
rather than every tool call.

OpenClaw's scanner-disagreement study~\cite{clawhubsignals} is this paper's premise
rather than a comparison point. Saha et al.\ demonstrate discovery- and
selection-manipulation attacks against ClawHub's ranking and 36.5--100\% governance
evasion against static scanners~\cite{skillmd2026}; their attacks target what a
human or ranking algorithm sees before install, complementary to our focus on what
an agent does after. Recent work on trajectory assurance argues individually
permitted actions can jointly violate a state-conditioned invariant that no deployed
mechanism enforces~\cite{trajectory2026}; our per-(resource, class) ledger is one
concrete instantiation of the enforcement point that work argues is missing, though
we enforce per-action rather than per-trajectory. Formal machinery for temporal
constraints over agent action sequences~\cite{temporal2026} is a natural direction
for extending $\rho$'s policy language beyond the per-action scope we implement.

\paragraph{Runtime governance as an emerging tier.} The argument that agent safety
requires a runtime control point is no longer only an academic position. A
commercial category has formed around it: venture-funded platforms now discover
agents across enterprise estates, record their tool invocations, and intervene on
anomalous behaviour in production, at reported estate sizes in the thousands of
agents. We take this as independent confirmation of the premise, and claim no part of
the runtime tier as our invention; several of these systems do things ours does not.
We do not cite a specific vendor deployment here, because we have not independently
verified any and would rather cite nothing than cite a figure we cannot check.

It also helps locate what we think is still open. In the published descriptions we
have read, of both registry-side scanning and this runtime tier, the decision offered
to an operator is binary: an action proceeds, or it raises an alert and waits for a
human. We have not found a published account of when scrutiny may be safely relaxed,
though we would not be surprised to learn that one exists or that such logic is
running in a product whose internals are not published. Absent it, the operator's
main recourse for reducing prompt volume is to turn the control down, which returns
to the difficulty above. \S\ref{sec:graduation} offers one derivation of how much
evidence a control should require before it stops asking. We put it forward as a
starting point rather than a settled answer, and \S\ref{sec:graduation} is candid
about what it does not give an operator.

\section{Conclusion}

OpenClaw's measurement showed that agent-skill security is not one scanner's
allow/block call, and we agree. We suggest one reason is that a publish-time pipeline
is built to compute maliciousness, while an operator also needs permission, which is
a different predicate evaluated against facts that do not exist at publish time. The
document those signals read is also only a partial bound on what runs: across 144
commands from 39 skills, 34.7\% of what the agent executed had a consequence class
absent from every code block of the document, and over 53 cleared skills that
document a forbidden action a live agent reached for one in 23.

We offer one control point for that gap: a deterministic resolver cheap enough
(67\,ms end to end, measured) to sit on every action, a per-lane trust ledger, and a derivation of
promotion thresholds from the operator's own stated tolerance rather than a vendor's
constant. We do not claim it is the only such design, or the best one, and we would
be glad to be improved on. \S\ref{sec:threat} is explicit about what this
implementation does not withstand, and \S\ref{sec:oats} about which parts of the
design are shipped and which are proposals.

We are equally direct about the limit. A skill whose harm is a hardcoded recipient,
an undisclosed scope, or a persuasive instruction produces no action for a gate to
resolve, and no amount of policy strictness changes that. That is precisely the
surface an artifact-level scanner reads and ours does not, which is the clearest
argument we can make for why these layers belong together rather than in competition.
Two candidate predicates that would have widened our reach failed our own
false-positive bar and were not shipped. The measurements point the same way: scan
the skill, and govern the action.

\section*{Availability}

The corpus study can be re-derived from public artifacts; the two live-agent
experiments cannot, and we set out which is which below rather than open with a claim
that covers only part of the paper. An earlier draft opened this section with ``every
measurement here can be re-derived from public artifacts,'' which was not true of
\S\ref{sec:divergence} or \S\ref{sec:runtime}. The corpus is
OpenClaw's own MIT-licensed dataset at
\url{https://huggingface.co/datasets/OpenClaw/clawhub-security-signals}; the
scanner-agreement figures quoted in \S\ref{sec:intro} are two SQL queries in its own
browser console. The profile, this paper, and the script reproducing
\S\ref{sec:permission} are at
\url{https://github.com/pheo-ai/open-agent-trust-system} under Apache 2.0. That
script runs the corpus through a released implementation rather than a private build,
so the result does not depend on access to ours:

\begin{lstlisting}
pip install pheo-oats pandas pyarrow requests
oats start --no-browser &
python research/reproduce_clawhub.py
\end{lstlisting}

This downloads roughly 1.6\,GB and runs for several hours over the full corpus. The
implementation used for every measurement here is \texttt{pheo-oats} 0.5.3, published
for macOS, Linux, and Windows at \url{https://pypi.org/project/pheo-oats/}. The
resolver's decision path contains no model, so a rerun reproduces the same classes
rather than approximating them.

\paragraph{What is not reproducible from this, and a disclosure.} Three limits, and
we would rather state them than have them found.

The resolver is not open source. \texttt{pheo-oats} is distributed under a
proprietary licence, and an auditor can re-execute it but cannot read the predicates
$\pi_k$. That is reproducibility by re-execution, not auditability, and it is a
weaker claim than \S\ref{sec:related}'s appeal to Anderson's third criterion invites.
We have published what can be published without the product: the full class set,
severity order, and graduation policy as \texttt{spec/action-classes.json}, which is
enough to check that a returned class is consistent with the stated ordering, and not
enough to check that a given command should have produced it.

The two live-agent studies now run from \texttt{research/live\_agent\_study.py},
released here, with every command they recorded in
\texttt{live\_agent\_divergence.jsonl} and \texttt{live\_agent\_runtime.jsonl}. The
container image is \texttt{research/sandbox/Dockerfile}. Both take
\texttt{--execute}, which is off by default: the corpus is unaudited publishers and
the actions under study fetch and run remote code, so pulling them should be a
deliberate choice. Without it the agent gets no output, spends its turns testing its
own tooling, and the study measures the harness; that is documented in the file,
because we ran it that way first.

What running it will not do is return these counts. A live agent is not
deterministic, and a different model on a different day will produce a different
trajectory. What we can offer is the method, the raw record of what we saw, and a
comparison against the earlier run of the same study: on the divergence figures the
two agree ($p = 0.78$), and on the runtime reach rate they do not
($p = 1.6 \times 10^{-5}$), for a reason \S\ref{sec:runtime} sets out. Treat the
numbers as one observation with its protocol attached, not as constants.

Pheo Inc.\ develops and sells the implementation evaluated here. This paper evaluates
our own product, against a bar we set, with no competing implementation measured
alongside it. \S\ref{sec:threat}'s benchmark is released partly to make that
comparison possible for someone other than us.

\end{document}